# High performance waveguide uni-travelling carrier photodiode grown by solid source molecular beam epitaxy

Xiaoli Lin,[1] Michele Natrella,[1] James Seddon,[1] Chris Graham,[1] Cyril C. Renaud,[1] Mingchu Tang,[1] Jiang Wu,[2] Huiyun Liu,[1] and Alwyn J. Seeds[1,*]

[1]*Department of Electronic and Electrical Engineering, University College London, London WC1E 7JE, UK*
[2]*University of Electronic Science and Technology of China, Chengdu, 611731, China*
*\*a.seeds@ucl.ac.uk*

**Abstract:** The first waveguide coupled phosphide-based UTC photodiodes grown by Solid Source Molecular Beam Epitaxy (SSMBE) are reported in this paper. Metal Organic Vapour Phase Epitaxy (MOVPE) and Gas Source MBE (GSMBE) have long been the predominant growth techniques for the production of high quality InGaAsP materials. The use of SSMBE overcomes the major issue associated with the unintentional diffusion of zinc in MOVPE and gives the benefit of the superior control provided by MBE growth techniques without the costs and the risks of handling toxic gases of GSMBE. The UTC epitaxial structure contains a 300 nm n-InP collection layer and a 300 nm $n^{++}$-InGaAsP waveguide layer. UTC-PDs integrated with Coplanar Waveguides (CPW) exhibit 3 dB bandwidth greater than 65 GHz and output RF power of 1.1 dBm at 100 GHz. We also demonstrate accurate prediction of the absolute level of power radiated by our antenna integrated UTCs, between 200 GHz and 260 GHz, using 3d full-wave modelling and taking the UTC-to-antenna impedance match into account. Further, we present the first optical 3d full-wave modelling of waveguide UTCs, which provides a detailed insight into the coupling between a lensed optical fibre and the UTC chip.

## 1. Introduction

TERAHERTZ (THz) band frequencies, located between microwaves and infrared in the electromagnetic spectrum (100 GHz to 10 THz), possess unique properties which can enable important potential applications in spectroscopy, material science, medical science, security imaging and high-speed wireless communication. To make these applications possible, efficient THz-wave generation and detection are required, with the former being the most challenging technology issue. Conventional electronic techniques, photonic techniques and quantum cascade lasers (QCLs) are important approaches for THz generation. With respect to frequency coverage and tunability, photonic techniques are considered superior to conventional electronic techniques [1]. QCLs can be very compact but are restricted to frequencies above 1.5 THz, need a cryogenic cooling system and the challenge is to raise the temperature of operation [2,3]. Photonic techniques can also benefit from signal transmission over optical fibre cables, allowing THz modulated optical signals to be distributed over long distances. Photodiodes convert light into electrical current and among various types of photomixing technologies [4], Uni-Travelling Carrier Photodiodes (UTC-PDs) have exhibited higher THz output power and larger bandwidth than competing structures. The generated low mobility holes do not travel across the collection layer. As majority carriers in the absorption layer, they respond within the dielectric relaxation time, which for p-$In_{0.53}Ga_{0.47}As$ $1\times10^{18}$ cm$^{-3}$ can be as short as 33 fs [5]. Therefore, only photogenerated electrons cross the collection layer and contribute to the UTC response [6]. As the velocity of electrons (3-5$\times10^7$ cm/s) is 6-10 times higher than that of holes ($5\times10^6$ cm/s) and electrons exhibit velocity overshoot in the carrier collection layer, UTC-PDs have a larger bandwidth than conventional PIN photodiodes. In addition, UTC-PDs generate higher output saturation current due to the reduced space charge effect in the depletion layer, which also results from the high electron velocity in the depletion layer [6]. These properties make UTC-PDs high speed and high output power photodiodes. Through continuous optimisation in semiconductor layer structures and improvement in thermal management, UTC-PD output power and bandwidth have been improved since the first devices were proposed in 1997 [5].

Until the mid-1990s, advanced phosphide epitaxy was only possible by Metal Organic Chemical Vapour Deposition (MOCVD) and gas-source Molecular Beam Epitaxy (GS-MBE) techniques, due to the problems associated with the high vapour pressure of white phosphorus, which is unfavourable for conventional Knudsen-type furnaces in solid-source MBE (SS-MBE) techniques [7]. In MOCVD and GS-MBE, arsine ($AsH_3$) and phosphine ($PH_3$) are commonly used as the source of arsenic and phosphorus. However, $AsH_3$ and $PH_3$ are not ideal because they are toxic (needing special and expensive safety precautions) and may introduce water vapour into the reactor chamber [7]. Solid-source MBE enables toxic gas free growth of phosphide compound semiconductors and can realise more precise doping profiles that would be difficult to achieve by MOCVD. To the best of the authors' knowledge, the first SS-MBE grown InGaAsP-based UTC-PD device was demonstrated in 2012 [8]. Fabricated vertically illuminated UTC-PDs exhibited 0.1 A/W responsivity at 1550 nm, 12.5 GHz 3 dB bandwidth and -5.8 dBm output power at 10 GHz, at a photocurrent of 4.8 mA. The work in [8] showed that it is possible to fabricate InGaAsP based UTC-PDs using materials grown by SS-MBE.

Though vertically illuminated UTC-PDs have shown improved 3 dB bandwidth, their geometry requires a stringent trade-off between DC responsivity and bandwidth. For such a normal-incidence structure, high responsivity requires a thick absorption layer, which results in increased carrier transit time and reduced device bandwidth. In waveguide UTC-PDs the trade-off between DC responsivity and bandwidth can be eased significantly. As light travels along the epitaxial layer direction, it is evanescently coupled from the passive waveguide into the absorption layer. As a result, for a given device area and epitaxial structure (i.e. same bandwidth), a waveguide coupled UTC-PD can exhibit higher output power than a vertically illuminated UTC-PD at high frequency [9, 10]. Output RF power of 10 dBm at 110GHz was achieved by the waveguide design, with DC responsivity of 0.36 A/W [9]. Waveguide coupled UTC-PDs are also compatible with the implementation of a Travelling-Wave design [11] which can substantially improve the device RC limited response. In a travelling wave structure, the optical signal travels at a group velocity comparable to the group velocity of the electrical signal. Travelling-Wave UTC-PDs (TW UTC-PDs) exhibited improved bandwidth and record level of THz figure of merit ($P_{THz} / P^2_{opt}$ in $W^{-1}$) [12, 13]. In [13] broadband and narrowband antennas were designed to couple TW UTC-PD THz output power to free-space. Resonant antenna integrated devices resulted in two narrowband emission peaks at 457 GHz and 914 GHz, achieving high emission power of 148 μW and 24 μW respectively, at a DC photocurrent of 13 mA. For devices integrated with bow-tie antennas, broadband emission was obtained, with 105 μW at 255 GHz and 30 μW at 408 GHz at a DC photocurrent of 13 mA.

In this paper, we discuss design, fabrication and characterisation of SS-MBE grown waveguide coupled UTC-PDs, integrated with antennas and coplanar waveguides. We compare measurement results with numerical results from 3d full-wave modelling at mm-wave frequencies. We also present a comprehensive optical 3d full-wave modelling analysis of the fibre-to-chip coupling.

## 2. Structure growth and device fabrication

The UTC-PD epitaxial structure grown by SS-MBE for this work is the same as the one we reported in [8], which was intended to achieve band gap engineering improvements (more accurately localised transitions) and more precise doping profiles than our previous work using MOCVD [13]. The detailed layer structure is shown in Table 1. The UTC epitaxial structure was grown on an Fe-doped SI (100) InP substrate at 490 °C in a Veeco Gen930 SS-MBE system [8], and Si and Be were used for n-type and p-type doping respectively. The 100 nm InP capping layer on top of the wafer was grown to protect the p-contact layer from oxidation. The 200 nm InGaAsP ($\lambda_g$ = 1.3 μm) layer functions as a diffusion block and P-contact. The 120 nm absorption layer consists of 5 levels of $In_{0.53}Ga_{0.47}As$ with graded doping concentrations, generating the quasi-electric field which accelerates electrons to higher velocity and helps them to be swept out from the layer. Two 10 nm InGaAsP ($\lambda_g$ = 1.3 μm and $\lambda_g$ = 1.1 μm respectively) spacer layers, between the absorption and the 300 nm n-InP collection layer, are used to "smooth" the abrupt conduction band barrier at the InGaAs-InP heterojunction interface. The 300 nm InGaAsP ($\lambda_g$ = 1.3 μm) waveguide layer was grown on the 700 nm $n^{++}$ - InP N-contact layer. To achieve the graded doping, the growth was stopped at each layer for 2 minutes to change the doping source temperature.

Table 1. UTC-PD Layer Structure Grown By SS-MBE

| \multicolumn{4}{c}{Solid Source MBE UTC-PD} | | | |
|---|---|---|---|
| **Doping** | **Material** | **Function** | **Thick (nm)** |
| >$1\times10^{19}$ | $p^{++}$ - $Q_{1.3}$ | p contact | 200 |
| $2.5 \times 10^{18}$ | $p^+$ - $In_{0.53}Ga_{0.47}As$ | absorber | 20 |
| $1\times10^{18}$ | $p^+$ - $In_{0.53}Ga_{0.47}As$ | absorber | 30 |
| $5 \times 10^{17}$ | p - $In_{0.53}Ga_{0.47}As$ | absorber | 30 |
| $2.5 \times 10^{17}$ | p - $In_{0.53}Ga_{0.47}As$ | absorber | 30 |
| u.i.d | u - $In_{0.53}Ga_{0.47}As$ | absorber | 10 |
| u.i.d | u - $Q_{1.3}$ | spacer | 10 |
| u.i.d | u - $Q_{1.1}$ | spacer | 10 |
| $1\times10^{16}$ | n - InP | collection | 300 |
| $2.5 \times 10^{18}$ | $n^+$ - $Q_{1.3}$ | waveguide | 300 |
| >$1\times10^{19}$ | $n^{++}$ - InP | n contact | 600 |
| S.I. Substrate | Fe Doped InP | substrate | 300 μm |

After removing the InP capping layer with HCl : $H_3PO_4$ (1 : 1), six mask patterns were used in sequence to fabricate the TW UTC-PD structures. Fabrication process steps included P-contact deposition, P-contact/absorption layer etching, waveguide etching, N-contact deposition, mesa etching, passivation and via etching, illustrated in Fig. 1. For P and N contacts, Ti/Pt/Au (75 nm/50 nm/400 nm) were sputtered using an SVS6000 and then processed by rapid thermal annealing (RTA) to form ohmic contacts with low contact resistivity ($\rho_c < 10^{-6}$ ohm $cm^2$). The fine device features were achieved by dry etch processing using Reactive Ion Etching (equipment: Plasma Pro NGP80). In the four rounds of dry etching (P-contact ridge etch, waveguide etch, mesa etch and vias etch), a soft mask (S1818) was used to form the etching pattern by UV photolithography and hard mask (SiN) was used to protect the device features during dry etching. Native oxide on the sidewall of the absorption, collection and waveguide layers (InGaAs, InP and InGaAsP respectively) contributes to surface leakage current and can increase the leakage current to several μA at -1 V reverse bias. In order to remove the unwanted oxide layer, 10% HCl solution was used for 1 min before P-contact and N-contact annealing and also before $SiO_xN_y$ insulation layer deposition.

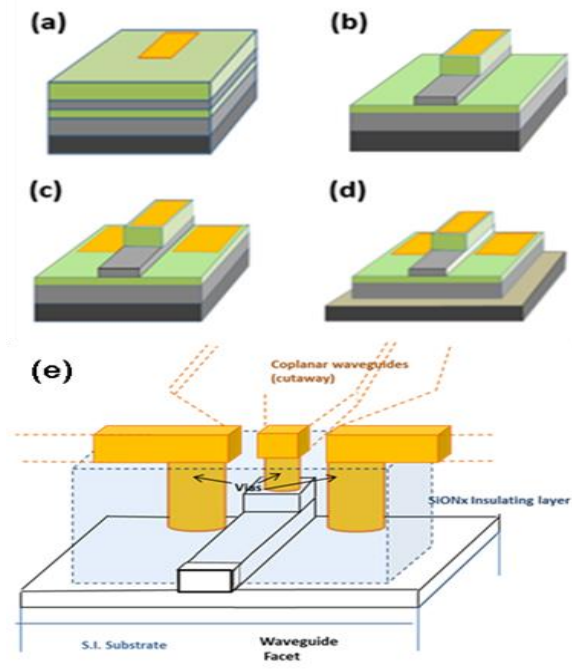

Fig. 1. Illustration of major fabrication steps: (a) P-contact deposition, (b) P-contact ridge etching and waveguide etching, (c) N-contact deposition, (d) mesa etching and (e) passivation and via etching.

## 3. Optical 3d full-wave modelling

In this section comprehensive optical 3d-full-wave modelling of waveguide UTC-PDs using CST transient solver is presented. A wide variety of algorithms have been developed for the simulation of passive photonic devices, though only a few have made it to mainstream use, such as the Beam Propagation Method (BPM), Eigenmode Expansion Method (EEM) and Finite Difference Time Domain (FDTD) [14].

The BPM method is an approximation technique for simulating the propagation of light in slowly varying optical waveguides; BPM struggles with handling structures containing material with significantly different refractive index, cannot deal with metals and reflections, and has problems modelling non-rectangular structures.

The EEM is a linear frequency-domain method which relies on the decomposition of the electromagnetic fields into a basis set of local eigenmodes that exist in the cross section of the device; although unable to deal with non-linearities, it provides a rigorous solution to Maxwell's equations, the main approximation being the number of modes used. It should be noted though, that the choice of the modes to be considered to simulate a device is still an assumption to be made by the user. However, for devices like the waveguide UTC-PDs, the type of propagation excited within the device by an optical beam incident externally, is one of the questions we want to investigate. The EEM cannot provide information on the external responsivity with respect to an incident Gaussian beam of arbitrary spot size. For the effect of the misalignment between incident beam and waveguide to be calculated using EEM it is necessary to add radiative modes.

The FDTD method is a finite-difference discretisation of Maxwell's equation in time and space and, in principle, can model virtually any structure, provided sufficient computing power is available. The Finite Integration technique (FIT) is a spatial discretisation scheme to solve electromagnetic field problems numerically in time and frequency domain. It was proposed in 1977 by Thomas Weiland [15] and has been continually developed since then [16]. It preserves basic topological properties of the continuous equations such as conservation of charge and energy. The FIT method applies Maxwell equations in integral form to a set of staggered grids and stands out due to high flexibility in geometric modelling and boundary handling as well as incorporation of arbitrary material distributions and material properties such as anisotropy, non-linearity and dispersion. Computing power has been increasing dramatically and becoming more accessible; these computational advances have particularly benefitted numerical techniques, such as FDTD and FIT, that discretise the entire computational domain, and thus larger and more complex scenarios can be investigated [16]. The CST Transient solver is based on the Finite Integration Technique (FIT). The FIT method covers the full frequency range of electromagnetics (from static up to high frequency) and optical applications and is the basis not only for the CST transient solver but also for other commercial simulation tools [16].

With the optical full-wave modelling performed in CST, we are able to simulate very realistic scenarios in which the excitation source is a Gaussian beam, which is a good approximation to the light coupled from a lensed optical fibre into the waveguide. In this way it is possible to calculate, among other things, the external responsivity, the effect of the misalignment between Gaussian beam and waveguide, the type of propagation excited within the waveguide (single- or multi-mode) and the way the light actually propagates and is absorbed throughout the UTC structure.

The material properties used to model the UTC-PD at 1550 nm (193.548 THz) are shown in Table 2. In particular, the gold metallisation was characterised with the gold optical model with dispersive permittivity, derived from [17].

**Table 2. Material optical properties at 1550 nm wavelength (193.548 THz)**

| Material | Relative permittivity | Tangent delta |
|---|---|---|
| InP | 10.11 | -- |
| $In_{0.53}Ga_{0.47}As$ | 12.88 | 0.0446 |
| InGaAsP | 11.49 | -- |

The optical analysis was carried out on a 3 x 15 $\mu m^2$ device. The geometrical details of the initial UTC optical model are illustrated in Fig. 2. This geometry is very similar to the UTCs fabricated for this work.

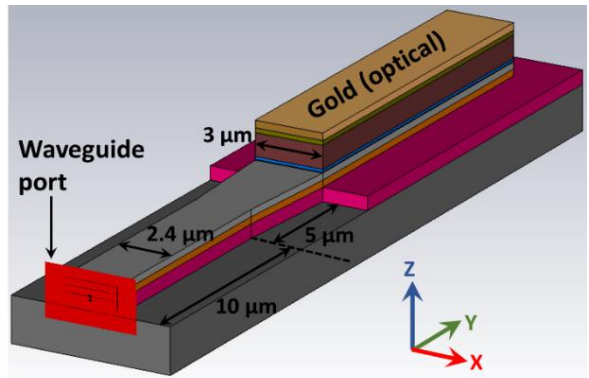

Fig. 2. Geometrical details of the initial UTC model used for the optical full-wave simulations.

For this first optical analysis a CST waveguide port was employed as excitation source. The waveguide ports represent a special kind of boundary condition of the calculation domain which enables the stimulation as well as the absorption of energy and the calculation of the waveguide modes. The electric field distribution, the power flow and absorption, generated by each mode, can be computed.

The power flow pattern along the Y direction and the propagation constant β of the two main modes are shown in Fig. 3. The optical waveguide does not support any transverse mode as both mode 1 and mode 2 are hybrid. The X component of the electric field is the dominant component for mode 1 which is similar to a $TE_{01}$ mode, while the X component of the magnetic field is the dominant component for mode 2 which is similar to a $TM_{01}$ mode.

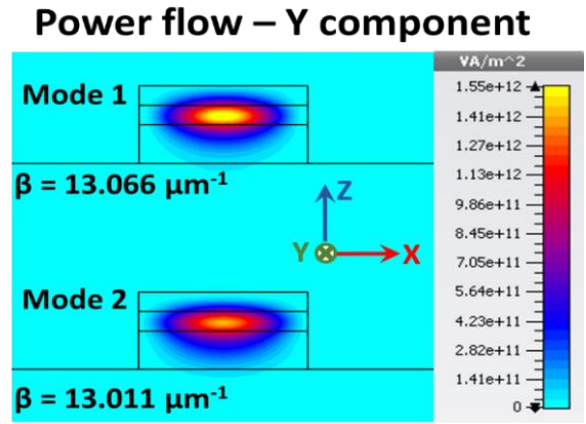

Fig. 3. Power flow pattern along the Y direction and the propagation constant β of the two main modes.

A side cross section of the power flow along the structure is given in Fig. 4, showing in particular the power coupling between the waveguide layer and the absorption layer in the ridge. Due to the single mode propagation, the intensity pattern does not change along the Y direction within the waveguide.

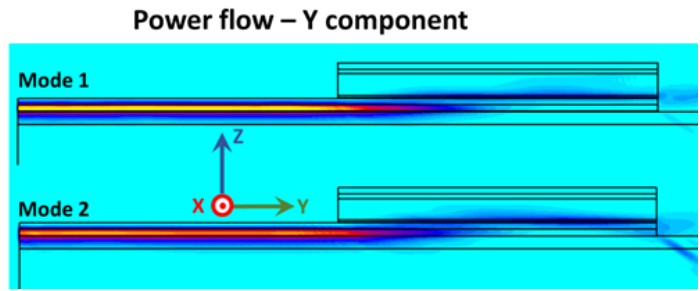

Fig. 4. Side cross section of the power flow along the structure, showing the power coupling between the waveguide layer and the absorption layer in the ridge.

In Fig. 5 a horizontal cross section cutting through the absorber is displayed, showing the way the light is absorbed along the layer. It is noted that the power stimulated by the waveguide port was equal to 0.5 W and that, at the frequency of interest (193.548 THz), the power coupled into the waveguide was virtually 100 %, i.e. 0.4997 W for mode 1 and 0.4984 W for mode 2. It follows that, by considering the ratio between the power absorbed in $In_{0.53}Ga_{0.47}As$ (0.3683 W for mode 1 and 0.2995 W for mode 2) and the power accepted, the internal quantum efficiency $\eta_{int}$, given in Fig. 5, can be calculated; the internal responsivity can also be evaluated as $\rho_{int} = \eta_{int}\, q / (hf)$ where q is the electron charge (1.602 x $10^{-19}$ C), h is the Plank constant (6.626 x $10^{-34}$ Js) and f is the frequency (193.548 THz).

The value of 0.921 A/W can hence be looked upon as the maximum theoretical responsivity attainable for the structure discussed in this section, ideally achievable if 100 % of the optical power coming out of the lensed fibre is coupled into the waveguide and propagates as mode 1. This scenario is far from being realisable because, as discussed in the next section, firstly only a fraction of the incident optical power will be successfully coupled into the waveguide and secondly because the waveguide concerned is multi-mode at 1550 nm free space wavelength. It is noted that when mode 1 and mode 2 were simulated simultaneously for the model in Fig. 2, the internal quantum efficiency was equal to 0.669, between that obtained for mode 1 and mode 2.

The power absorption pattern in Fig. 5 suggests that the single modes travelling through the input waveguide excite multimode propagation in the waveguide represented by the UTC ridge structure.

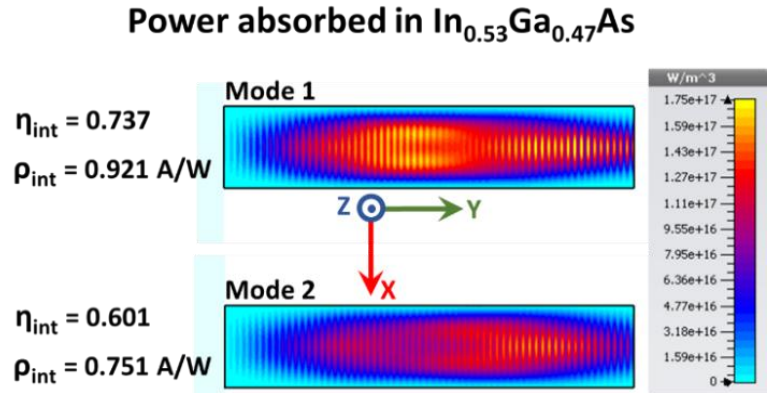

Fig. 5. Horizontal cross section through the absorber showing details of the power absorbed in the In$_{0.53}$Ga$_{0.47}$As layer for mode 1 and mode 2.

To assess the importance of the waveguide length, the simulation for mode 1 was repeated for a number of longer waveguides and no significant difference was observed. The power coupled in the waveguide is well confined and propagates with negligible losses. Furthermore, because of the single mode propagation, the power flow along the propagation direction remains constant and the intensity pattern arriving at the ridge is the same, regardless of the waveguide length; as a consequence, the power coupling between the waveguide and the absorption layer always occurs in the same fashion. The internal quantum efficiency and the pattern of the power absorbed throughout the In$_{0.53}$Ga$_{0.47}$As are also the same for different lengths. For real devices the waveguide geometry is not as perfect as in our model, therefore some power will be scattered and lost along the optical waveguide; also, as already mentioned, single mode propagation is hard to achieve in real optical waveguides.

Now a more realistic scenario is analysed, in which a Gaussian beam is the source of the optical power. Two linearly polarised (X and Z) Gaussian beams, with nominal electric field amplitude of 1 V/m and spot size of 4 µm, were used to couple light into the UTC optical waveguide. The X polarised beam was intended to excite the propagation of mode 1, whose dominant component is the electric field X component, while the Z polarised beam was intended to excite mode 2, for which the magnetic field X component is dominant and the electric field is principally in the Z direction. Details of the beam for the X polarisation case are given in Fig. 6. The figure shows the propagating electric field in free space, though the device structure is outlined in order to discuss the alignment between beam and waveguide. The Gaussian beam wave front is planar at the waist, therefore, for perfect alignment, the facet of the cleaved waveguide needs to lie within the same plane; also, the beam optical axis must pass through the centre of the waveguide layer cleaved facet. As we set the nominal electric field amplitude of two inputting

Gaussian beams to 1 V/m, the corresponding incident optical power was 8.2563 x 10$^{-15}$ W for the X polarised beam and 8.4285 x 10$^{-15}$ W for the Z polarised beam. No anti-reflection coating was included in the model as this was not applied to the UTCs fabricated for this work.

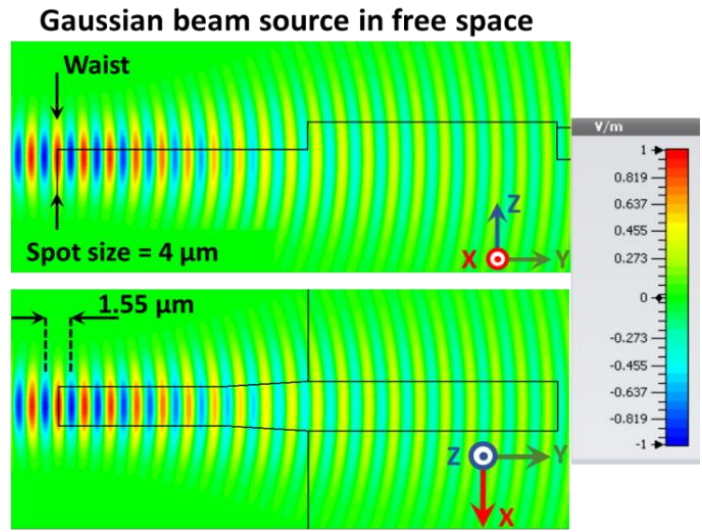

Fig. 6. Gaussian beam details for the X polarisation case. The figure shows the propagating electric field in free space, though the device structure is outlined in order to highlight the alignment between beam and waveguide.

A side cross section view and a horizontal cross section view (through the waveguide layer) of the power flow along the Y direction, are shown in Fig. 7 and Fig. 8 respectively, for both X and Z polarisation. The propagation excited in the waveguide from the X polarised and Z polarised Gaussian beam is different, however it is clearly multimode in both cases.

The incident optical power of the Gaussian beam in our model represents well the optical power measured with the power meter, at the output of the lensed fibre in the real laboratory experiment. As discussed previously, the power lost in the absorption layer is also provided as part of the simulation results. It is then possible to calculate the external quantum efficiency and particularly the external responsivity, which is the quantity actually measured experimentally.

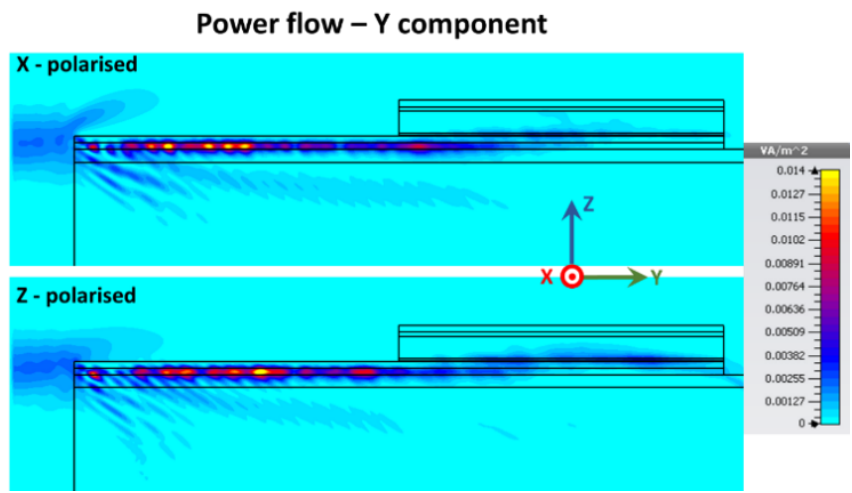

Fig. 7. Side cross section view of the power flow along the Y direction.

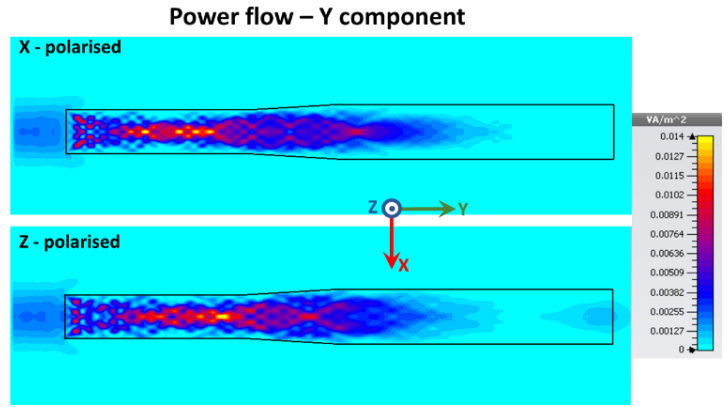

Fig. 8. Horizontal cross section view (through the waveguide layer) of the power flow along the Y direction.

The external quantum efficiency and responsivity are given in Fig. 9 for the X and Z polarisation, together with a horizontal cross section of the absorption layer showing the power absorption pattern. The Z polarised beam provides a slightly higher quantum efficiency despite mode 1 (whose main electric field component is along the X direction) having exhibited a superior performance in the single mode analysis; this may be due to the fact that the results shown in Fig. 9 actually originate from a multimode propagation and/or to the fact that the fraction of incident power coupled into the waveguide may be higher for the Z polarised beam.

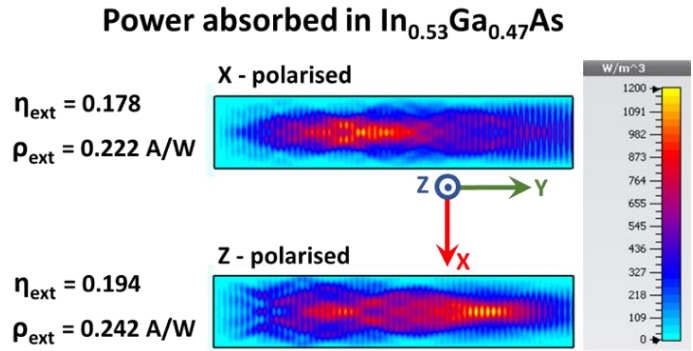

Fig. 9. Horizontal cross section, cutting through the absorber, showing details of the power absorbed in the $In_{0.53}Ga_{0.47}As$ layer for both the X and Z polarised input Gaussian beams. The external quantum efficiency and responsivity are also given.

The highest experimental values of responsivity we measured (i.e. 0.20 A/W, 0.21 A/W and 0.22 A/W) were very close to the numerical values of responsivity shown in Fig. 9.

As in the simulations using the waveguide ports, discussed above, we investigated the influence of the waveguide length. For this analysis, only the X polarised Gaussian beam was employed. Fig. 10 shows the power flow on a horizontal cross section through the waveguide layer, for three different values of waveguide length, i.e. 15 µm, 35 µm and 50 µm. In real waveguides, due to fabrication imperfections, some power is scattered and lost as the light travels along the waveguide and therefore UTCs with longer waveguides will exhibit lower optical responsivity. The waveguides discussed in this section are ideal and have no losses, therefore longer waveguides show a self-imaging effect, as can be seen in Fig. 10. Nevertheless, unlike the single mode propagation case, the intensity pattern in the multimode case is not constant over the propagation direction, because each mode travels with a different velocity and the way the modes combine varies along Y. For this reason, the intensity pattern arriving at the UTC ridge changes with the input waveguide length and it is interesting to see how this affects the device responsivity.

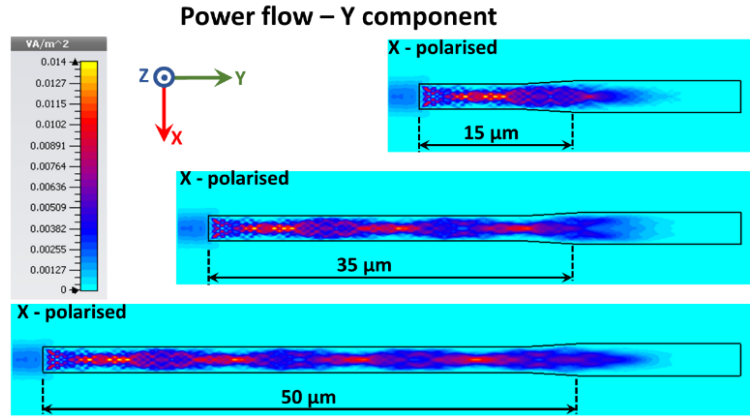

Fig. 10. Power flow on a horizontal cross section through the waveguide layer, for three different values of waveguide length, i.e. 15 µm, 35 µm and 50 µm.

The external quantum efficiency, the external responsivity and the power absorption pattern in the $In_{0.53}Ga_{0.47}As$ layer, are given in Fig. 11, for the three different waveguide lengths.

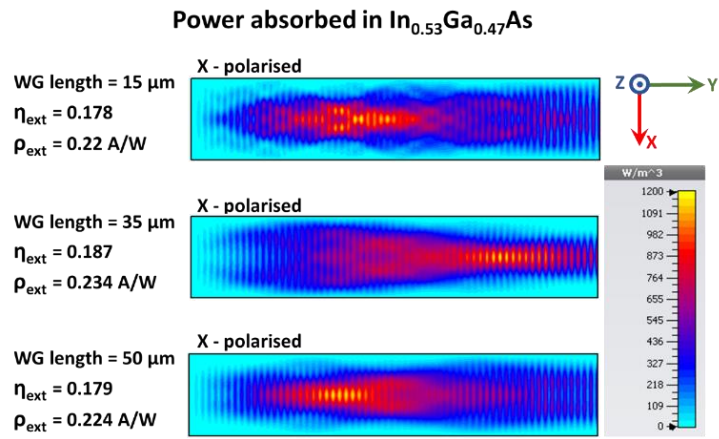

Fig. 11. Quantum efficiency, external responsivity and power absorption pattern in the $In_{0.53}Ga_{0.47}As$ layer, for the three different waveguide lengths.

The power absorption patterns in the three different cases are all different, unlike the single mode case in which the pattern was not dependent on the waveguide length. Unlike the single mode case, the external quantum efficiency varies in a perceptible way, e.g. + 5 % for the 35 µm waveguide case with respect to the 15 µm long waveguide. The fact that the quantum efficiency variation with respect to the waveguide length is not monotonic, supports the conclusion that such variations are not due to power escaping the waveguide during the propagation, but rather to the varying intensity profile.

Lastly, an additional aspect of the fibre to chip coupling, which can be investigated by means of the 3d full wave modelling, is the effect of misalignment between Gaussian beam and optical waveguide. The misalignment for the case of the X polarised Gaussian beam described in Fig. 6 and the waveguide UTC structure in Fig. 2 is discussed. Misalignments along all the three axes are considered and indicated as ΔX, ΔY and ΔZ. The misalignments are considered with respect to the perfect alignment, realised when the beam optical axis is perpendicular to waveguide facet, passes through the centre of the waveguide layer facet and the beam waist is located at the same Y coordinate as the waveguide facet. The beam position is fixed and the UTC waveguide structure is translated by ΔX, ΔY or ΔZ; for instance ΔX = -750 nm means that the chip is moved by 750 nm along the X axis in the negative direction.

The responsivity versus the misalignment is plotted in Fig. 12. The misalignment along Z seems to be the most sensitive, as expected, since the waveguide layer is 2.4 µm wide (X direction) but only 300 nm thick (Z direction); a thicker waveguide layer would be desirable, also because it would lessen the chance of exciting substrate modes while coupling the light from the lensed fibre, however it is challenging to grow InGaAsP layers thicker than 300 nm by solid source molecular beam epitaxy.

It is noted that the case Δz = 540 nm corresponds with the Gaussian beam optical axis passing through the centre of the absorber facet. The result confirms that, even for a very short waveguide, the optical power coupled directly through free space is not relevant and the optical responsivity decreases monotonically as the negative displacement along Z increases.

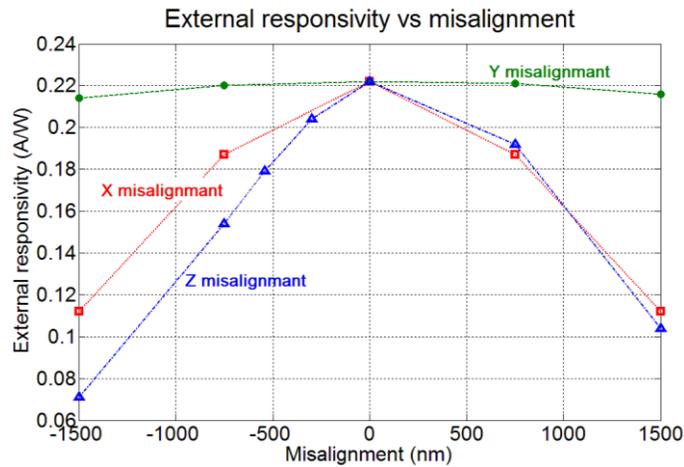

Fig. 12. External responsivity vs misalignment.

## 4. Device characterisation

### 4.1 Coplanar waveguide coupled UTC-PDs

After 2 µm thick silicon oxynitride ($SiO_xN_y$) was deposited to cover the whole UTC-PD as a passivation and insulation layer, Ti/Au coplanar waveguides (CPW) were sputtered on top of the device to allow RF power to be extracted and measured by means of air coplanar ground-signal-ground probes. Vias were etched through SiOxNy down to P contact layer and N-contact layer before CPW deposition. Fig. 13 shows the schematic diagram of the CPW UTC-PD (a) and the photo of a fabricated CPW UTC-PD bar with 2.5 µm spot size lensed fibre placed in front of the waveguide (b). The light coming from the lensed fibre is coupled into UTC-PD through the passive waveguide. The length of the waveguide depicted in the schematic diagram in Fig. 13 (a) is shorter than in reality.

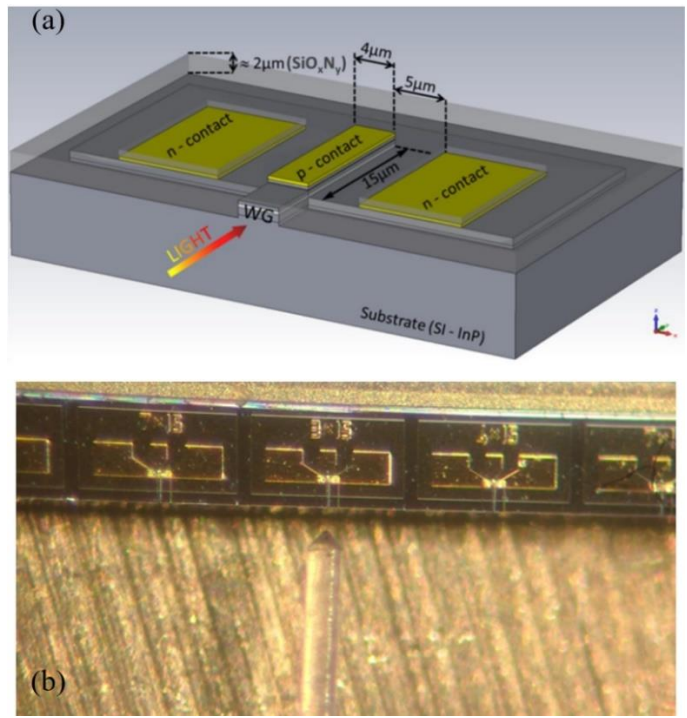

Fig. 13. (a) Schematic diagram of the CPW UTC-PD structure, (b) fabricated CPW UTC-PD bar containing different p-contact size UTC-PDs (from left to right: 7x15 μm$^2$, 3x15 μm$^2$ and 4x15 μm$^2$).

Typical fabricated CPW integrated UTC-PDs have exhibited dark currents of less than 1 μA at −1 V reverse bias, see Fig. 14. As shown in Fig. 15 (a) and (b), responsivities up to 0.19 A/W and 0.22 A/W were measured for 4 μm × 15 μm and 7 μm × 15 μm devices respectively, at -2V bias voltage and without applying anti-reflection coating on the facet of the waveguide. The responsivities are improved with respect to that reported (0.1 A/W) for normal incidence illuminated UTC-PDs grown by SS-MBE, which had a 20 μm diameter optical window [8]. The waveguide coupled UTC-PDs in this paper have the same epitaxial structure and layers thickness as the vertical illuminated UTC-PD reported in [8]. In waveguide UTC-PDs, the InGaAsP waveguide layer beneath the collection layer, enables evanescent coupling with the absorption layer, allowing for higher saturation power because it increases the length over which the absorption takes place and also increases the responsivity of the UTC-PD [18]. Compared with the evanescently-coupled waveguide UTC-PD fabricated using MOCVD, our SS-MBE grown UTC-PDs showed higher responsivity. MOCVD grown 70 μm$^2$ waveguide MUTC-PDs reported in [19] exhibited 0.15 A/W responsivity at 1550 nm wavelength. The responsivity of SS-MBE grown waveguide UTC-PD (4 μm × 15 μm) was measured up to 0.19 A/W at 1550 nm.

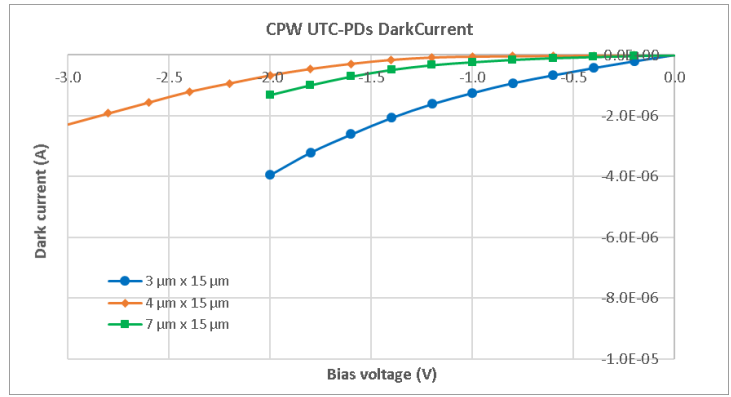

Fig. 14. Dark current of CPW UTC-PDs with treatment by 10% HCl (1 min) before $SiO_xN_y$ passivation layer deposition.

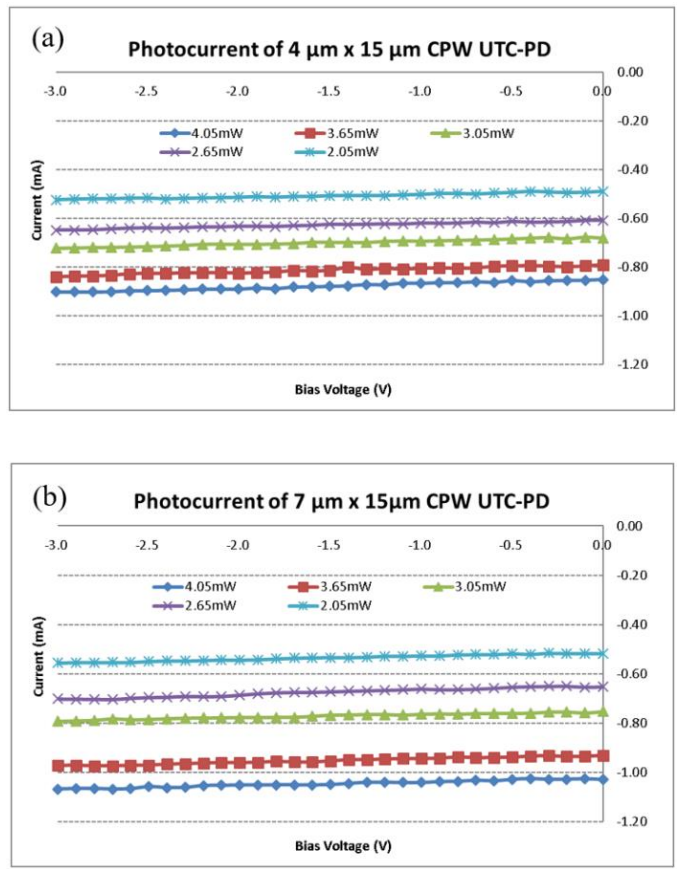

Fig. 15. Measured photocurrent of CPW UTC-PDs with different input optical power at bias voltage 0 V to -3 V. Responsivities for 4 μm × 15 μm and 7 μm × 15 μm devices at -2V are 0.20 A/W and 0.22 A/W, respectively.

The device frequency response, measured by the Lightwave Component Analyser (calibrated to open, short and 50 Ω load), is plotted in Fig. 16. The devices were biased with -2V voltage and the input 1550nm optical power was 5 dBm. As shown in Fig. 16, the 7 μm × 15 μm and 4 μm x 15 μm devices exhibited 3 dB bandwidth of 47 GHz and 65 GHz respectively, while the 3 μm × 10 μm device exhibited bandwidth greater than 67 GHz. It is obvious to see the improvement in 3dB bandwidth with respect to the 12.5 GHz 3dB bandwidth of our previous work on vertical illuminated UTC-PDs [8]. The 3dB bandwidth improvement is due to lower capacitance (because of the smaller size device) and lower series resistance.

Our 3 µm × 10 µm waveguide UTC-PD, has higher 3dB bandwidth (> 70 GHz) than the waveguide UTC-PD in [20] (55 GHz), despite the device in [20] has a smaller size (1.5 µm × 10 µm) and thinner absorption layer (100 nm). The 3dB bandwidth is enhanced by the graded doping in the absorption layer. The graded doping in the p-type UTC absorption layer yields a potential gradient, and hence a quasi-field, which effectively accelerates electrons in the direction of decreasing doping levels, i.e. from the absorption towards the collection layer.

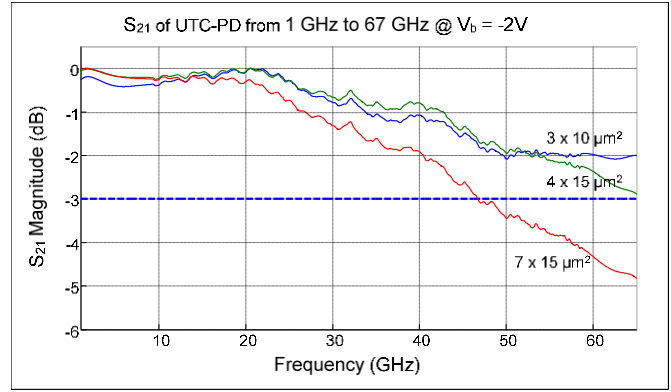

Fig. 16. $S_{21}$ response from 1 GHz to 67 GHz of CPW coupled UTC-PDs.

The output power of CPW coupled waveguide UTC-PDs was extracted with W-band (75–110 GHz) coplanar probe coupled to an Agilent E4418B EPM series power meter via a Flann flexible W-band waveguide. The good agreement on power measurement between spectrum analyser and EPM series power meter is illustrated in [12]. The experimental arrangement for CPW UTC-PD power measurement is shown in Fig. 17. The laser tones from laser 1 and laser 2 were adjusted to generate an optical heterodyne frequency of 100 GHz. Two polarization controllers were used on each laser output to set the laser polarisations to the same state. The optical tones are then combined in a 50:50 coupler and amplified using an Erbium Doped Fibre Amplifier (EDFA). An Amplified Spontaneous Emission (ASE) filter was placed after the EDFA to reduce the contribution from ASE to the photocurrent. The optical beat signal was coupled to the UTC-PD waveguide via a lensed optical fibre with a 2.5 µm spot size. A polarization controller was included before the lensed fibre to optimize the coupling of light to the waveguide. Reverse bias was applied to the device via the integrated bias tee in the coplanar probe. Throughout the measurements, the device was cooled using a Peltier cooler set to 20 ºC.

The RF Power and average photocurrent dependence on incident optical power is plotted as a function of reverse bias voltage in Figure 18. The maximum RF power obtained increases with reverse bias voltage. The maximum RF power recorded was 1.1 dBm with a photocurrent of 13.1 mA and reverse bias of 3 V.

The maximum measured output power in this case is higher than the recently reported 106 GHz output power in [21], that is -3.8 dBm biased at -2 V and in [22] where an output power of -7 dBm is reported at 100 GHz for a waveguide integrated p-i-n photodiode. Comparable powers were measured from our previous state-of-the-art UTC-PDs in [9] (0.8 dBm output power at 110 GHz at 13 mA photocurrent and -2V bias). For the device evaluated in this experiment, thermal management was impeded by a substrate thickness of 350 µm. Reduced substrate thickness enabled photocurrents up to 36 mA to be used in [9], yielding a record output power of 10 dBm. High powers at 100 GHz have also been obtained using narrow-band matching circuits [23] or flip chip bonding [24] of a vertically illuminated Modified UTC-PD (MUTC-PD) with an inductive peaking circuit have been incorporated into the device design. RF Powers of 13.2 dBm and 9 dBm for the respective devices were recorded. However, the higher emitted power comes at the expense of reduced bandwidth. For waveguide integrated MUTC-PDs, a maximum power of 5.1 dBm at 120 GHz is reported in [25], for a device with 0.5 A/W responsivity and saturation photocurrent of 30 mA. This is still lower than the maximum values reported in [9] for a waveguide integrated UTC-PD.

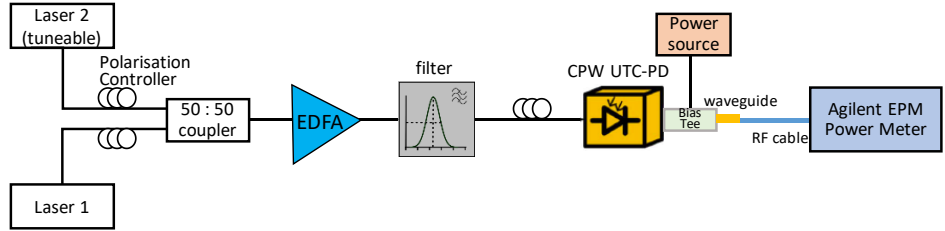

Fig. 17. CPW coupled UTC-PD output power measurement arrangement.

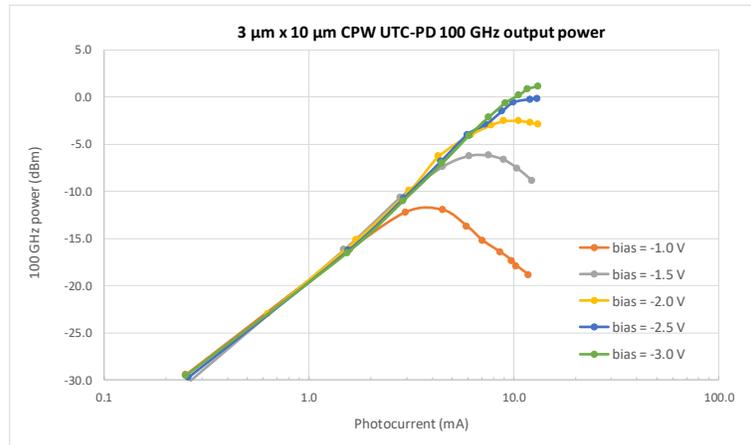

Fig. 18. RF Output power vs photocurrent of a 3 μm × 10 μm CPW UTC-PD at 100 GHz as a function of bias.

Based on previous experience reducing the substrate thickness for the devices under test [9] or using flip-chip mounting on diamond or aluminum nitride [26] could be used to increase thermally limited power output further. It is unclear from the present measurements, however whether the saturation at high optical powers is due to thermal or space charge effects and a more detailed study of the thermal properties at high optical power is needed.

### 4.2 Antenna integrated waveguide UTC-PDs

The antenna structure of Fig. 19 (b) was deposited on the $SiO_xN_y$ insulation layer. The deposited antenna was connected to the UTC P-contact and N-contact through via holes as small as 2 μm. The 2 μm fine features were achieved by RIE processing. The antenna impedance had been designed to realise complex-conjugate matching to the UTC-PD in [27] at a frequency of 250 GHz. The total power emitted by an antenna depends on the impedance match with the driving source and the radiation efficiency. The absolute level of power emitted by an antenna over the frequency range can be calculated and predicted if the UTC impedance is known and the coupling efficiency between UTC and antenna taken into account [27] [28]. We presented a comprehensive study of UTC-PD output impedance and frequency photo-response in [27]. The use of a Silicon lens combined with antenna integrated UTC-PDs is an established solution to couple THz power into free space [29]. Our design of antenna integrated UTC-PD assembled with a Si lens is shown in Fig. 19 (a). The far field radiation pattern and directivity at 250 GHz, calculated with 3d full-wave modelling, is shown in Fig. 20. The modelling of the whole structure including the 6 mm diameter Si lens, reveals that part of the electric field is trapped within the lens by reflections and generates strong standing waves, as shown in Fig. 21. In future work this issue can be addressed by applying an index matching layer to the lens surface.

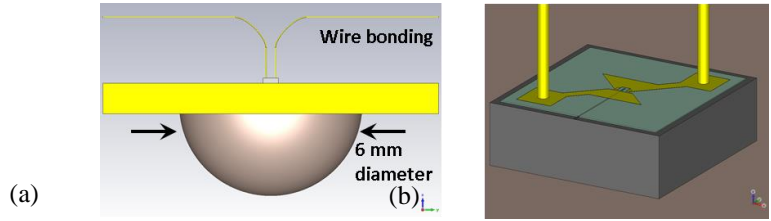

Fig. 19. Wire-bonded antenna integrated UTC-PD on a Si lens (D = 6 mm).

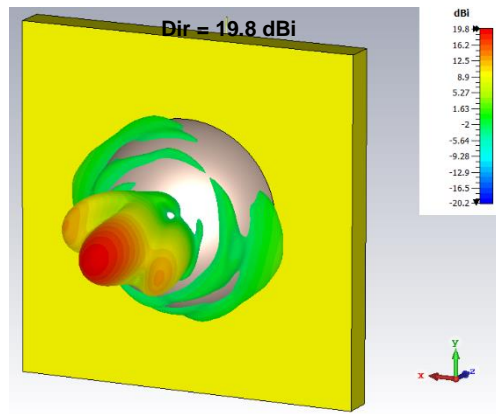

Fig. 20. Far field radiation pattern and directivity at 250 GHz.

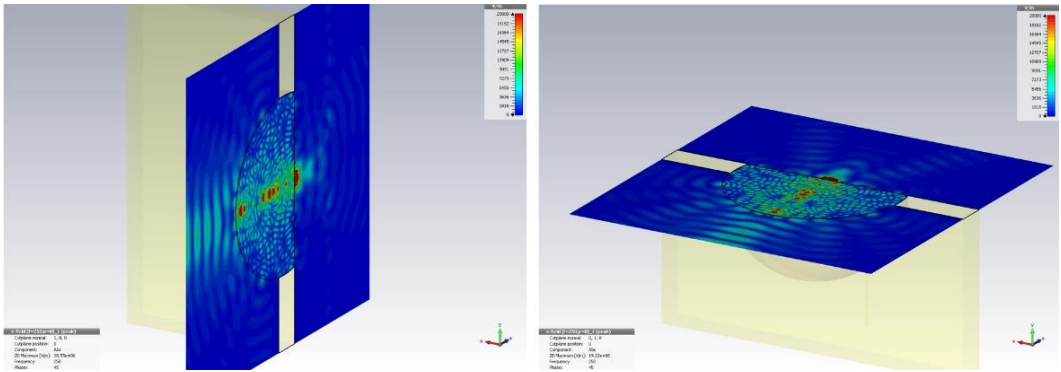

Fig. 21. Cross section views of the electric field magnitude at 250 GHz for an antenna integrated UTC-PD integrated with a 6mm diameter Si lens.

By means of 3d full-wave modelling and using the knowledge of UTC impedance [27] [28], we have calculated the absolute level of power emitted by our antenna integrated UTCs with Si lenses, as shown in Fig. 22, for 10 mA and 13.5 mA photocurrent. The power radiated by our fabricated antenna-integrated

UTC-PD was measured with a Thomas Keating opto-acoustic power sensor from 200 GHz to 260 GHz. The test device had a P-contact ridge size of 4 μm × 15 μm, and it was mounted on a 6 mm diameter Si lens and wire-bonded. The experimental arrangement is shown in Fig. 24. The measured UTC leakage current was below 100 nA at -1V bias and the responsivity was 0.14 A/W. At -2.6 V DC bias, the measured radiated powers at 250 GHz were 32 μW and 60 uW, with photocurrent of 10 mA and 13.5 mA respectively. The measured radiated powers (200 GHz to 260 GHz) plotted in Fig. 22 show good agreement with the numerical results. The radiated RF power is relatively low because the bow-tie antenna was not impedance matched with the UTC-PD, as the antenna had been designed to be impedance matched with the UTC-PDs of [27]. The radiated powers calculated for the case of the antenna driven by the impedance matched UTCs in [27], are plotted in Fig. 23, for photocurrents of 10 mA and 20 mA. For 10 mA photocurrent, the calculated RF power is 124 μW at 250 GHz, 5 times higher than the calculated power plotted in Fig. 22. This modelled radiation power of waveguide UTC-PDs is slightly higher than the reported emission power of the UTC-PDs in [13] (105 μW at 255 GHz). For the non-resonant bow-tie antenna integrated with vertically illuminated UTC-PD, the highest reported output power at 280 GHz is around 16 μW at 6 mA photocurrent [30]. Our bow-tie antenna integrated UTC-PD has similar output power at this frequency and is expected to generate higher radiation power when impedance matched to the antenna.

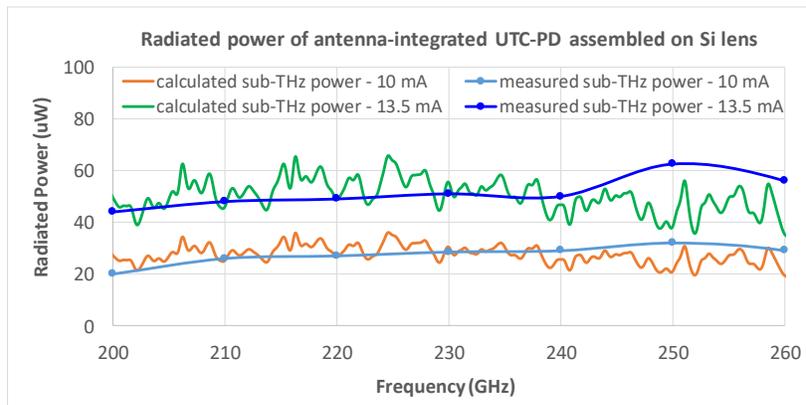

Fig. 22. Power radiated by antenna integrated UTC-PDs (3 × 15 um$^2$) mounted on a 6 mm diameter Si lens at bias of -2V. Here the antenna is not impedance matched with the UTC-PDs.

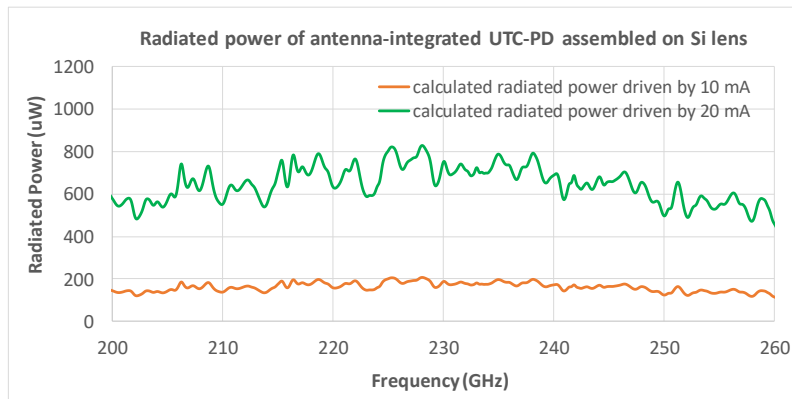

Fig. 23. Calculated power radiated by antenna integrated UTC-PDs (3 × 15 um$^2$) mounted on a 6 mm diameter Si lens, with antenna impedance matched with UTC-PD impedance.

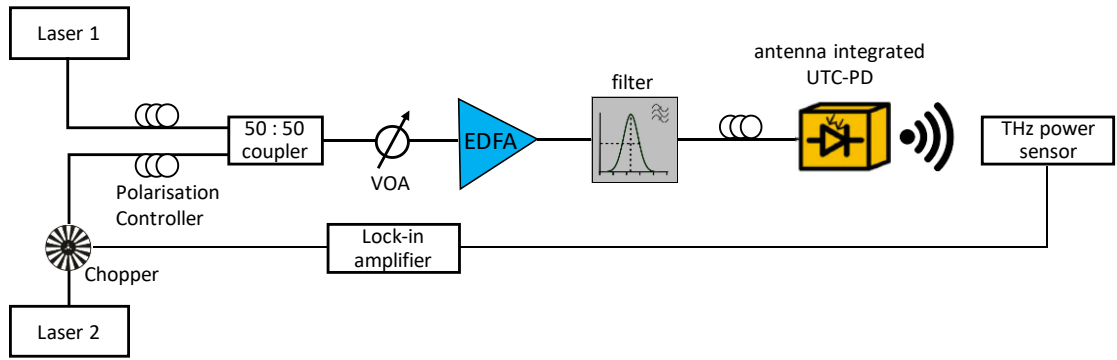

Fig. 24. Radiated power measurement arrangement. VOA indicates a variable optical attenuator.

## 5. Conclusion

We have reported the first waveguide UTC-PDs fabricated on InGaAsP-based epitaxial structures grown by SS-MBE, to combine the merits of InGaAsP materials for UTC-PD performance with the advantages of SS-MBE. CPW integrated UTC-PDs and antenna integrated UTC-PDs were fabricated and characterised. CPW integrated UTC-PDs of 4 μm × 15 μm and 7 μm × 15 μm area, exhibited responsivities of 0.19 A/W and 0.22 A/W respectively. The 3 μm × 15 μm devices achieved 3 dB bandwidth exceeding 65 GHz at -2V DC bias, showing a substantial improvement on previously reported vertically illuminated UTC-PDs grown by SS-MBE. Output RF power greater than 1 dBm was obtained at a frequency of 100 GHz for an average photocurrent of 13 mA. We have also demonstrated accurate prediction of the absolute level of radiated power, between 200 GHz and 260 GHz. Finally, we show the first optical 3d full wave modelling of waveguide UTCs, providing detailed insight into the coupling between a lensed optical fibre and the waveguide UTC chip.

## 6. Acknowledgement

The authors would like to acknowledge financial support from the UK Engineering and Physical Sciences Research Council through the HyperTHz Programme Grant (EP/P021859/1).

**References**

1. T. Nagatsuma, H. Ito, and T. Ishibashi, "High-power RF photodiodes and their applications" Laser Photonics 3, 123–137 (2009).
2. R. Köhler, A. Tredicucci, F. Beltram, H. E. Beere, E. H. Linfield, A. G. Davies, D. A. Ritchie, R.C. Iotti, and F.Rossi, "Terahertz semiconductor heterostructure laser," Nature 417, 156–159 (2002).
3. M. A. Belkin, Q. J. Wang, C. Pflügl, A. Belyanin, S. P. Khanna, A. G. Davies, E. H. Linfield, and F. Capasso, "High-Temperature operation of terahertz quantum cascade laser sources," IEEE J. Sel. Topics Quantum Electron 15(3), 952–967( 2009).
4. D. Saeedkia and S. Safavi-Naeini, "Terahertz photonics: Optoelectronic techniques for generation and detection of terahertz waves," J. Lightw. Technol. 26(15), 2409-2423 (2008).
5. Tadao Ishibashi, Satoshi Kodama, Naofumi Shimizu and Tomofumi Furuta, "High-Speed Response of Uni-Traveling-Carrier Photodiodes", Jpn. J. Appl. Phys. 36, 6263-6268 (1997).
6. T. Nagatsuma, H. Ito, and T. Ishibashi, "High-power RF photodiodes and their applications," Laser Photonics Rev., 3(1–2), 123–137 (2009).

7. M. Pessa, M. Toivonen, M. Jalonen, P. Savolainen, and A. Salokatve, "All-solid-source molecular beam epitaxy for growth of III–V compound semiconductors," Thin Solid Films 306(2), 237-243 (1997).
8. Michele Natrella, Efthymios Rouvalis, Chin-Pang Liu, Huiyun Liu, Cyril C. Renaud, and Alwyn J. Seeds, "InGaAsP-based uni-travelling carrier photodiode structure grown by solid source molecular beam epitaxy", Optics Express, 20(17), 19279-19288 (2012).
9. C. Renaud, D. Moodie, M. Robertson, and A. Seeds, "High output power at 110 GHz with a waveguide uni-travelling carrier photodiode," The 20th Annual Meeting of the IEEE, 782–783 (2007).
10. C. Renaud, M. Robertson, D. Rogers, R. Firth, P. Cannard, R. Moore, and A. Seeds, "A high responsivity, broadband waveguide uni-travelling carrier photodiode," Proc. SPIE, 6194(1), C1940–C1940 (2006).
11. K. S. Giboney, M. J. W. Rodwell, and J. E. Bowers, "Traveling-Wave photodetector design and measurements," IEEE J. Sel. TopicsQuantum Electron., 2(3), 622-629 (1996).
12. E. Rouvalis, C. Renaud, D. Moodie, M. Robertson, and A. Seeds, "Continuous wave terahertz generation from ultra-fast InP-based photodiodes," IEEE Trans. Microw. Theory Tech. 60(3), 509–517 (2012).
13. E. Rouvalis, C. C. Renaud, D. G. Moodie, M. J. Robertson, and A. J. Seeds, "Traveling-wave uni-traveling carrier photodiodes for continuous wave THz generation," Optics Express 18(11), 11105–11110 (2010).
14. D. Gallagher, "Photonic CAD Matures," IEEE LEOS Newsl., 22(1), 8–14 (2008).
15. T. Weiland, "A discretization model for the solution of Maxwell's equations for six-component fields," Archiv Elektronik und Uebertragungstechnik, Vol. 31, 116–120, (1977).
16. C. Warren, S. Sesnic, A. Ventura, L. Pajewski, D. Poljak, and A. Giannopoulos, "Comparison of Time-Domain Finite-Difference, Finite-Integration, and Integral-Equation Methods for Dipole Radiation in Half-Space Environments," Prog. Electromagn. Res. M **57**, 175–183 (2017).
17. P. B. Johnson and R. W. Christy, "Optical constants of the noble metals," Phys. Rev. B, 6(12), 4370–4379 (1972).
18. Alwyn Seeds, Cyril Renaud, Michael Robertson, "Photodetector Including Multiple Waveguides", U.S. Patent 7851782 B2, Dec. 14 (2010).
19. Qinglong Li, Keye Sun, Kejia Li, Qianhuan Yu, Patrick Runge, Willi Ebert, Andreas Beling, Joe C. Campbell, "High-power waveguide MUTC photodiode with 70 GHz bandwidth", Microwave Photonics (MWP), 2016 IEEE International Topical Meeting, 225-228, (2016).
20. Y. Muramoto, K. Kato, M. Mitsuhara, O. Nakajima, Y. Matsuoka, N. Shimizu and T. Ishibashi, "High-output-voltage, high speed, high efficiency uni-travelling-carrier waveguide photodiode", Electronics Letters, 34(1), 122-123 (1998).
21. Toshimasa Umezawa, Atsushi Kanno, Kouichi Akahane,Atsushi, Matsumoto, Naokatsu Yamamoto, et al, "Study of high power generation in UTC-PD at 110-210 GHz", Proc. SPIE 10531, Terahertz, RF, Millimeter, and Submillimeter-Wave Technology and Applications XI, 1053115, (2018).
22. S. Nellen, R. Kohlhaas, L. Liebermeister, S. Breuer, B. Globisch, M. Schell, "Continuous Wave Terahertz Generation from Photodiode-Based Emitters with up to 200 μW Terahertz Power", 43rd International Conference on Infrared, Millimeter, and Terahertz Waves (IRMMW-THz), (2018).
23. H. Ito, T. Nagatsuma, A. Hirata, T. Minotani, A. Sasaki, Y. Hirota, T. Ishibashi, "High-power photonic millimetre wave generation at 100 GHz using matching-circuit-integrated uni-travelling-carrier photodiodes", IEE Proceedings – Optoelectronics, 150(2), 138-142 (2003).
24. Andreas Beling, Jesse S. Morgan, Keye Sun, Qianhuan Yu, "High Power Integrated 100 GHz Photodetectors", International Topical Meeting on Microwave Photonics (MWP), (2018).
25. Gan Zhou, Patrick Runge, Shahram Keyvaninia, Sten Seifert, Willi Ebert, Sven Mutscha, "High-Power InP-Based Waveguide Integrated Modified Uni-Traveling-Carrier Photodiodes", Journal of Lightwave Technology 35(4), 717 – 721 (2017).
26. Xiaojun Xie, Qiugui Zhou, Kejia Li, Yang Shen, Qinglong Li, Zhanyu Yang, Andreas Beling, and Joe C. Campbell, "Improved power conversion efficiency in high-performance photodiodes by flip-chip bonding on diamond", Optica, 1(6), 429-435 (2014).
27. M. Natrella, C.-P. Liu, C. Graham, F. van Dijk, H. Liu, C. C. Renaud, and A. J. Seeds, "Accurate equivalent circuit model for millimetre-wave UTC photodiodes," Optics Express, 24(5), 4698–4713 (2016).
28. Michele Natrella, Chin-Pang Liu, Chris Graham, Frederic van Dijk, Huiyun Liu, Cyril C. Renaud, and Alwyn J. Seeds, "Modelling and measurement of the absolute level of power radiated by antenna integrated THz UTC photodiodes," Opt. Express, 24(11), 11793-11807 (2016).
29. J. Van Rudd and D. M. Mittleman, "Influence of substrate-lens design in terahertz time-domain spectroscopy," JOSA B, 19(2), 319–329 (2002).
30. Hiroshi Ito, Tadao Ishibashi, "Terahertz-wave generation using resonant-antenna-integrated uni-traveling-carrier photodiodes", Proc. SPIE, Image Sensing Technologies: Materials, Devices, Systems, and Applications IV, 102090R, (2017).